\documentclass[12pt]{article}

\usepackage{amssymb}
\usepackage{amsmath}
\usepackage{bm}
\usepackage{upgreek}
\usepackage{graphicx}

\begin{document}

\begin{center}
{\bf CASIMIR FRICTION AT ZERO AND FINITE TEMPERATURES}

\vspace{1cm}
Johan S. H{\o}ye\footnote{johan.hoye@ntnu.no}

\bigskip
Department of Physics, Norwegian University of Science and Technology, N-7491 Trondheim, Norway

\bigskip
Iver Brevik\footnote{iver.h.brevik@ntnu.no}

\bigskip
Department of Energy and Process Engineering, Norwegian University of Science and Technology, N-7491 Trondheim, Norway

\bigskip
\today
\end{center}

\begin{abstract}
The Casimir friction problem for dielectric plates that move parallell to each other is treated by assuming one of the plates to be at rest. The other performs a closed loop motion in the longitudinal direction. Therewith by use of energy dissipation the formalism becomes more manageable and transparent than in the conventional setting where uniform sliding motion is assumed from $t=-\infty$ to $t=+\infty$. One avoids separating off a reversible interparticle force (independent of friction) from the total force. Moreover, the cases of temperatures $T=0$ and finite $T$ are treated on the same footing. For metal plates we find the friction force to be proportional to $v^3$ at $T=0$ while at finite $T$ it is proportional to $v$ for small $v$ as found earlier. Comparisons with earlier results of Pendry (1997, 2010) and Barton (2011) are made.

\end{abstract}


\section{Introduction}

The Casimir friction force experienced between two parallel semi-infinite dielectric plates moving longitudinally with respect to each other at micro{\-}meter or nanometer separation is a topic that has recently attracted considerable interest. This may actually appear somewhat surprising, in view of the expected smallness of the effect, presumably beyond direct measurability in practice. However, the problem is one of fundamental interest as it implies, as a dissipative phenomenon, the interaction between a particle system and a heat reservoir. It is even conceivable that it is related to the dynamic Casimir effect. Some references by others dealing with Casimir friction are \cite{teodorovich78,pendry97,pendry98,pendry10,volokitin99,volokitin03,volokitin07,volokitin08,volokitin11,dedkov08,dedkov10,
dedkov11,dedkov12,philbin09,barton10A,barton10B,barton11,barton11B,silveirinha13A,maghrebi13}.

In this work we will evaluate the friction force via the energy dissipated. We assume that the  lower plate (2) is at rest, while the upper plate (1) executes motion in a {\it closed loop}, which means that it finally slides back to its starting position. By final return to the stating position there will be no work associated with a reversible force. Both plates are taken to be nonmagnetic and consist of the same material, with  permittivity $\varepsilon(\omega)$. The specification of the motion is as follows:

Without loss in generality we can assume that the closed loop motion takes place in the $x$ direction. Initially, at $t<-\tau-\alpha \tau$ where  $\tau \gg 1$, $\alpha \gg1$, we assume that the upper plate is at rest. In the period $t \in (-\tau-\alpha \tau, -\tau)$, it moves to the left with a low constant velocity $-v/\alpha$ ($v$ is constant and positive). Then, for $t\in (-\tau, +\tau)$, it moves to the right with finite velocity $v$, whereafter it moves to the left again  with low velocity $-v/\alpha$ until it returns to its initial position at   $t=\tau +\alpha \tau$. In this way the motion of the upper plate upper plate executes a closed loop.

The reason to use piecewise constant velocity is to facilitate evaluation of integral (\ref{26}) below, especially in the non-linear regime for large $v$. Then the limit $\alpha\rightarrow\infty$ is taken such that the velocity as well as the corresponding dissipation outside the interval $-\tau<0<\tau$ approaches zero. (But the region outside this interval can not be disregarded. Otherwise the velocity from the starting position and back to it indirectly will be infinite, i.e.~$\alpha\rightarrow 0$.)

\bigskip

The use of energy dissipation makes it possible to consider a more general situation and thus implies certain advantages:

\begin{itemize}

\item One avoids linearization of the interaction around some relative position as done in our previous papers. Thus more general motions can be considered.
    \item One avoids separating a reversible part of the interparticle force from the total force. The reversible force depends upon the position but does not depend upon the velocity. So it does not change sign when the velocity $v$ changes sign, and is thus not associated with a friction force.
        \item
        The formalism treats the cases of $T=0$ and of finite $T$ or a combination of them on the same footing. Thus we can derive the friction force proportional to $ v^3$ at $T=0$ (first found by Pendry \cite{pendry97}) and proportional to $ v$ for small $v$ at finite $T$ within the same mathematical approach.
\end{itemize}

We make use of the same statistical mechanical approach as spelled out in our earlier recent papers \cite{hoye13,hoye10,hoye11A,hoye11B,hoye12A,hoye12B}; cf. also the earlier papers \cite{hoye92,brevik88}. This approach consists essentially in applying  the Kubo formula to time-dependent cases. The extension of our approach to energy dissipation was initiated in Ref.~\cite{hoye10} for a pair of particles in slow relative motion. There the relative motion was described by the position $x(t)=vte^{-\eta t}$ for $t>0$ for its closed loop motion ($x(0)=x(\infty)=0$) for small $\eta\rightarrow 0$. For this situation with small velocity we verified that the energy dissipation $\sim\int(\dot x(t))^2\,dt$ led to the same result as the friction force times velocity integrated.

\bigskip
In the next section we establish the general formalism according to the closed-loop model, and derive expressions for the dissipated energy and the friction force per unit surface area. In Sec.~\ref{sec3} we reestablish our previous results in linear theory valid when $v$ is low. Sec.~\ref{sec4} is devoted to a discussion of the $T=0$ case. And in Sec.~\ref{sec5} comparisons with results obtained in Refs.~\cite{pendry97}, \cite{pendry10}, \cite{volokitin07}, and \cite{barton11B} are made.

\section{Energy dissipation}
\label{sec2}

Consider a quantum mechanical two-oscillator system whose reference state is one of uncoupled motion corresponding to a Hamiltonian $H_0$. The equilibrium situation is then perturbed by a time-dependent term written in general form as $-AF(t)$ where $A$ is a time independent operator and $F(t)$ is a classical function that depends upon time $t$. Thus the Hamiltonian becomes $H=H_0-AF(t)$. For simplicity consider first a pair of one-dimensional oscillators for which we can write \cite{hoye10}
\begin{equation}
-AF(t)=\psi({\bf r}(t))s_1 s_2
\label{1}
\end{equation}
where $\psi({\bf r})$ is the coupling strength. The ${\bf r}$ is the separation between the oscillators, and $s_1$, $s_2$ are the vibrational coordinates of the oscillators.

If the oscillators move with relative velocity ${\bf v}$ the ${\bf r}(t)={\bf r}_0+{\bf v}t$. Then in Ref.~\cite{hoye10} the $\psi$ was expanded around ${\bf r}(t)={\bf r}_0$ for small ${\bf v}$ and a friction linear in ${\bf v}$ was found. This force vanished at temperature $T=0$. The reason for this lack of friction is that linearization means ${\bf v}\rightarrow 0$
and thus slow variations in interaction by which the oscillators can not be excited out of their ground states to absorb or dissipate energy. However, for higher velocities and thus larger displacements the varying interaction will feel non-zero frequencies, and it can not be expanded to linear oder in ${\bf v}$. Also it is not clear how to separate out the friction part of the resulting force from the reversible part of it. To circumvent this latter problem we will instead evaluate the energy dissipated from which the friction force for a given velocity can be obtained \cite{hoye10,hoye11A,hoye12A}. Equivalence with time-dependent quantum mechanical perturbation theory was shown in Refs.~\cite{hoye11A} and  \cite{hoye11B}.

With interaction (\ref{1}) the force between the oscillators is
\begin{equation}
{\bf B}=-\nabla\psi({\bf r})s_1 s_2.
\end{equation}
The thermal average of this force from the perturbing interaction is given by the Kubo formula \cite{hoye10,brevik88,kubo59,landau80}
\begin{equation}
\langle {\bf B(t)}\rangle=\int\limits_{-\infty}^t\phi_{BA}(t-t')F(t')\,dt'
\label{3}
\end{equation}
where the response function $\phi_{BA}$ (here vector since ${\bf B}$ is vector) is given by
\begin{equation}
\phi_{BA}(t)=\frac{1}{i\hbar}{\rm Tr}\{\rho[A,{\bf B}(t)]\}.
\label{4}
\end{equation}
Here $\rho$ is the density matrix and ${\bf B}(t)$ is the Heisenberg operator ${\bf B}(t)=e^{itH/\hbar}{\bf B}e^{-itH/\hbar}$ where the $s_1 s_2$ part of ${\bf B}$ like $A$ is time independent. Now we put
\begin{equation}
F(t)=-\psi({\bf r}(t))
\label{5}
\end{equation}
such that $A=s_1 s_2$, and we can write
\begin{eqnarray}
\nonumber
\phi_{BA}(t)&=&\nabla\psi \,\phi(t)\\
\phi(t)&=&{\rm Tr}\{\rho C(t)\}, \quad C(t)=\frac{1}{i\hbar}[s_1 s_2, s_1(t) s_2(t)].
\label{6}
\end{eqnarray}
With electrostatic interaction the $\psi({\bf r})$ is given by the Coulomb interaction (\ref{34}) below where it further on is replaced with the dipole-dipole interaction. But it should be noted that by our approach the interaction is not limited to the electromagnetic one.

Dissipation of energy is obtained by multiplying the average force (\ref{3}) with velocity and the integrate. However, the system has to return to its starting position to eliminate the reversible force. We write the position as
\begin{equation}
{\bf r}(t)={\bf r}_0+{\bf v} q(t)
\label{7}
\end{equation}
where ${\bf r}_0$ is the relative initial position between the particles. This should also be their final position, i.e.~$q(\pm\infty)=0$).
Their relative velocity is ${\bf v}\dot q(t)$.

Note that unlike we did earlier, we here will not linearize interaction (\ref{5}) around the position ${\bf r}_0$ since more arbitrary motion not limited to small velocities will be considered. Earlier we used $\psi({\bf r}(t))=\psi({\bf r}_0)+{\bf v}q(t)\nabla\psi({\bf r}_0)+\cdots$ where the $\nabla\psi$ term generated the friction force. In the present approach the $\nabla$ by the Fourier transform method which is used below, reappears as $-i{\bf k}_\perp$ in the exponential of the quantity $A(t,t')$ in Eq.~(\ref{11}). (The previous linear regime is recovered by linearization of the exponential.)

  With Eq.~(\ref{5}) we thus have
\begin{equation}
\dot F(t)=-({\bf v}\nabla\psi)\dot q(t).
\label{8}
\end{equation}
Altogether the dissipated energy becomes
\begin{equation}
\Delta E({\bf r}_0)=-\int\limits_{-\infty}^\infty {\bf v}\dot q(t)\langle{\bf B}\rangle\,dt=-\int\limits_{-\infty}^\infty\int\limits_{-\infty}^t \dot F(t)\phi(t-t')F(t')\,dt'\,dt.
\label{9}
\end{equation}
With somewhat different reasoning the same expression for $\Delta E$ with $F(t)=q(t)$ was established in earlier works \cite{hoye10,hoye11A,hoye12A}.

Now let two half-planes slide parallel to each other. They are filled with oscillators with uniform number densities $\rho_1$ and $\rho_2$ respectively. For low densities the resulting energy dissipation is the sum of contributions from all particle pairs. In the continuum limit this energy dissipation per unit surface area is then with ${\bf r}_0=\{x_1, y_1, z_1-z_2\}$ ($x_2=y_2=0$)
\begin{equation}
\Delta E=\rho_1 \rho_2\int\limits_{z_1>d,z_2<0} \Delta E({\bf r}_0)\,dx_1dy_1dz_1dz_2
\label{10}
\end{equation}
where $d$ is the separation between the planes parallel to the $xy$-plane. With this the origin is located at the surface of the lower  plane.

The ${\bf r}_0$-dependence of Eq.~(\ref{9}) is present in the $F(t)$ so we can write
\begin{equation}
\Delta E=\rho_1 \rho_2\int L(t,t')\phi(t-t')\,dtdt'
\label{10a}
\end{equation}
So by use of Fourier transform methods in the $x$- and $y$-directions we find with expression (\ref{7}) inserted ($\nabla\rightarrow-i{\bf k}_{\perp}$, $d{\bf k}_\perp=dk_xdk_y$, ${\bf v}||{\bf k}_\perp$, $z_0=z_1-z_2$)
\begin{eqnarray}
\nonumber
L(t,t')&=&-\int \dot F(t)F(t')\,dx_1dy_1dz_1dz_2\\
&=&-\frac{1}{(2\pi)^2}\int\limits_{z_1>d,z_2<0}\hat\psi(z_0,{\bf k}_\perp)\hat\psi(z_0,-{\bf k}_\perp)A(t,t')\,d{\bf k}_\perp dz_1dz_2
\label{11}
\end{eqnarray}
with $A(t,t')=-i{\bf k}_\perp {\bf v}\dot q(t) e^{-i{\bf k}_\perp {\bf v}(q(t)-q(t'))}$. This follows from the Fourier transform
\begin{eqnarray}
\nonumber
&&  \int\psi({\bf r}(t))\exp(i{\bf k_\perp}{\bf r}_{0\perp})\,dx_0 dy_0\\
\nonumber
&=&\exp{(-i{\bf k}_\perp {\bf v}q(t))}
\int\psi({\bf r}(t))\exp(i{\bf k_\perp}{\bf r}_{\perp})\,dx dy\\
&=&\hat\psi(z_0,{\bf k}_\perp)\exp{(-i{\bf k}_\perp {\bf v}q(t))}
\label{11a}
\end{eqnarray}
where ${\bf r}_{0\perp}=\{x_0,y_0\}$, ${\bf r}_\perp=\{x,y\}$. With the Coulomb interaction the $\hat\psi(z_0,{\bf k}_\perp)$ is given by Eq.~(\ref{35}) below which further on is replaced by the corresponding dipole-dipole interaction (\ref{35b}).

The $A(t,t')$ is to be integrated together with the $\phi(t-t')$ of Eqs.~(\ref{9}) and (\ref{10}). Thus we first write
\begin{equation}
A(t,t')=\dot Q(t,\omega_v)(Q(t',-\omega_v)+1)
\label{12}
\end{equation}
where
\begin{equation}
Q(t,\omega_v)=e^{-i\omega_v q(t)}-1,\quad \mbox{with}\quad \omega_v={\bf k}_\perp {\bf v}.
\label{13}
\end{equation}
Without changing the dissipated energy this can first be modified into
\begin{equation}
A(t,t')=\dot Q(t,\omega_v)Q(t',-\omega_v).
\label{14}
\end{equation}
The reason for this is that when integrating $\dot Q(t,\omega_v)$ alone together with $\phi(t-t')$ (in Eq.~(\ref{9})) the result is zero since $\phi$ depends only upon the difference $t-t'$. Thus integration first with respect to $t'$ gives a constant, and the second integration of $\dot Q$ with respect to $t$ then gives zero since we have assumed $q(t)\rightarrow 0$ as $t\rightarrow\pm\infty$ (requirement of return to starting position). Further the $A$ can be symmetrized with respect to $\omega_v$ without changing integral (\ref{11}). So finally
\begin{equation}
A(t,t')=\frac{1}{2}\sum\limits_{n=\pm 1} \dot Q(t,n\omega_v)Q(t',-n\omega_v).
\label{15}
\end{equation}
For the $t$-integrations the $\phi(t-t')$ is needed. An explicit expression for $\phi(t)$ was found in Ref.~\cite{hoye10}. (In Ref.~\cite{hoye11A} it was expressed as a sum of matrix elements while in Ref.~\cite{hoye12A} its Fourier transform was used.) Thus from Eqs.~(14) and (15) of Ref.~\cite{hoye10} we have
\begin{equation}
\phi(t)=D[(2\langle n_1 \rangle+1)\cos(\omega_1t)\sin(\omega_2 t)
+(2\langle n_2 \rangle+1)\cos{(\omega_2t)}\sin(\omega_1 t)]
\label{16}
\end{equation}
with
\begin{equation}
D=\frac{\hbar}{2m_1m_2\omega_1\omega_2},\quad 2\langle n_i \rangle+1=\coth{(\frac{1}{2}\beta\hbar\omega_i)}, \quad (i=1,2)
\label{17}
\end{equation}
with $\beta=1/(k_BT)$ where $k_B$ is Boltzmann's constant. The expression for $D$ is for oscillators where the coordinate   $s_x=x$ is length. When $s$ is an oscillating dipole moment the $s_x=qx$ and $\alpha=q^2/(m\omega^2)$ is polarizability where here the $q$ is charge. Then with Eq.~(\ref{6}) for $\phi(t)$ the $D$ is multiplied with $q^4$ to have
\begin{equation}
D=\frac{1}{2}\hbar\omega_1\omega_2\alpha_1\alpha_2.
\label{18}
\end{equation}

Combining Eqs.~(\ref{16}) and (\ref{17}) the response function $\phi(t)$ can be written as ($t>0$)
\begin{eqnarray}
\nonumber
\phi(t)&=&C_-\sin\omega_-t+C_+\sin\omega_+t\\
C_\pm&=&\frac{H}{\hbar}\sinh\left(\frac{1}{2}\beta\hbar\omega_\pm\right)
\label{19}
\end{eqnarray}
with $\omega_\pm=|\omega_1\pm\omega_2|$ ($\phi(t)=0$ for $t<0$) and
\begin{equation}
H=\frac{\hbar^2\omega_1\omega_2\alpha_1\alpha_2}{4\sinh(\frac{1}{2}\beta\hbar\omega_1)\sinh(\frac{1}{2}\beta\hbar\omega_2)}.
\label{20}
\end{equation}
Now the $A(t,t')$ can be integrated with $\phi(t-t')$. This has the same form as the integral performed in Sec.~3 of ref.~\cite{hoye11A} where for small ${\bf v}$ (linear case) one has $Q(t,\omega_v)=i\omega_v q(t)+\cdots$.

To evaluate the dissipated energy in more detail we write Eqs.~(\ref{10}) - (\ref{11}) as
\begin{equation}
\Delta E=\frac{\rho_1 \rho_2}{(2\pi)^2}\int\limits_{z_1>d, z_2<0}\hat\psi(z_0,{\bf k}_\perp)\hat\psi(z_0,-{\bf k}_\perp)J(\omega_v)\,d{\bf k}_\perp dz_1 dz_2.
\label{24}
\end{equation}
where
\begin{equation}
J(\omega_v)=-\int\int\limits_{t>t'}A(t,t')\phi(t-t')\,dtdt'=C_- I(\omega_-)+C_+ I(\omega_+).
\label{23}
\end{equation}
With expression (\ref{19}) for $\phi(t)\rightarrow\phi(t-t')$ and  $\sin\varphi=(e^{i\varphi}-e^{-i\varphi)}/(2i)$ ($\varphi=\omega t-\omega t'$) we get for $I(\omega)$ ($\omega=\omega_\pm$) the integral
\begin{equation}
\label{21}
I(\omega)=-\frac{1}{2i}\int\int\limits_{t>t'}A(t,t')(e^{i\omega t}e^{-i\omega t'}-e^{-i\omega t}e^{i\omega t'})\,dtdt'.
\end{equation}
The $A$ is given by Eq.~(\ref{15}). By partial integration and then exchange of integration variables in the last term (with $n\rightarrow -n$) this becomes ($Q(t,n\omega_v)\rightarrow 0$, $t\rightarrow\pm\infty$)
\begin{eqnarray}
\nonumber
I(\omega)&=&\frac{\omega}{2}\int\limits_{-\infty}^\infty\int\limits_{-\infty}^\infty\frac{1}{2}\sum\limits_{n=\pm 1}Q(t,n\omega_v)Q(t',-n\omega_v)e^{i\omega t}e^{-i\omega t'}\,dtdt'\\
&=&\frac{\omega}{4}\sum\limits_{n=\pm 1}\hat Q(-\omega,n\omega_v)\hat Q(\omega,-n\omega_v)
\label{22}
\end{eqnarray}
with $\hat Q(\omega,-n\omega_v)=\int Q(t,-n\omega_v)e^{-i\omega t}\,dt$. Here the two terms of Eq.~(\ref{21}) have become one term where the condition $t>t'$ on the limits of integration has vanished.

To obtain the dissipation the $q(t)$ is to  be specified; cf. the Introduction.  Constant velocity ${\bf v}$ is  chosen between times $-\tau$ and $\tau$.  The system moves very slowly from its starting position at time $t=-(\alpha+1)\tau$ and returns slowly to it at $t=(\alpha+1)\tau$. In the limit $\alpha\rightarrow\infty$ the latter slow motion will not contribute to dissipation. Thus

\begin{equation}
q(t)=\left\{
\begin{array}{cccc}
&-\tau&-\displaystyle{\frac{t+\tau}{\alpha}},\quad & -(\alpha+1)\tau<t<-\tau\\
&t,& &-\tau<t<\tau\\
&\tau&-\displaystyle{\frac{t-\tau}{\alpha}},\quad & \tau<t<(\alpha+1)\tau.
\end{array} \right.
\label{25}
\end{equation}
By evaluation of the Fourier transform one finds
\begin{eqnarray}
\nonumber
\hat Q(\omega,-\omega_v)&=&\int\limits_{-\infty}^\infty(e^{i\omega_v q(t)}-1)e^{-i\omega t}\,dt\\
\nonumber
&=&2\left[\frac{(1+1/\alpha)\omega_v\sin((\omega-\omega_v)\tau)}{(\omega+\omega_v/\alpha)(\omega-\omega_v)}-\frac{(\omega/\alpha)\sin(\omega(1+\alpha)\tau)}{(\omega+\omega_v/\alpha)\omega}\right]\\
&\underset{\alpha\rightarrow\infty}{\longrightarrow}&
2\frac{\omega_v\sin((\omega-\omega_v)\tau)}{\omega(\omega-\omega_v)}
\underset{\omega_v\rightarrow0}{\longrightarrow}
2\omega_v\frac{\sin(\omega\tau)}{\omega^2}.
\label{26}
\end{eqnarray}
The requirement to return to the initial position can be noticed from Eq.~(\ref{26}). Motion limited to the interval $-\tau<t<\tau$ corresponds to  letting $\alpha\rightarrow\infty$ by which $\hat Q(\omega,-\omega_v)\rightarrow\sin((\omega-\omega_v)\tau)/(\omega-\omega_v)-\sin(\omega \tau)/(\omega)$. As start and finish positions should be the same, they indirectly would imply infinite starting and finishing velocities, and evaluation of the sought dissipation would no longer be meaningful.
Finally, we are interested in the limit $\tau\rightarrow\infty$ by which $\hat Q^2$ becomes a $\delta$-function. Its amplitude is determined by the integral $\int_{-\infty}^\infty(\sin x/x)^2\,dx=\pi$, by which expression (\ref{22}) for $I(\omega)$ becomes ($\omega=\omega_\pm$)
\begin{equation}
I(\omega)=\pi\tau\frac{\omega_v^2}{\omega}[\delta(\omega-\omega_v)+\delta(\omega+\omega_v)].
\label{27}
\end{equation}
Eq.~(\ref{27}) when inserted in Eq.~(\ref{23}) will determine the dissipated energy (\ref{24}) from which the friction force per unit area is obtained ($\tau\rightarrow\infty$, $v=v_x$)
\begin{equation}
F=-\frac{\Delta E}{2\tau v}.
\label{28}
\end{equation}

Eq.~(\ref{27}) may be commented in view of the arguments and derivations made in Refs.~\cite{pendry97} and \cite{volokitin07}. There the friction force is regarded as a momentum transfer  due to the Doppler shift when photons are transmitted between the plates. However, the photons are virtual and connected to the evanescent fields close to the surfaces. With $\omega=\omega_1-\omega_2$ it will be reasonable to interprete the $\delta$-function argument $\omega\pm\omega_v$ as a Doppler shift, i.e.~$\omega_1=\omega_2\mp\omega_v$. The viewpoint and methods used for our evaluations in this work, however, are different. Below we will use the electrostatic interaction where no photons are involved. (For short separation this should be equivalent to the evanescent field.) Instead of Doppler shift of photons the $\omega_v={\bf k}_\perp{\bf v}$ are here regarded as the frequencies produced due to the varying interaction when the particles move relative to each other. These frequencies generate the quantum mechanical transitions within the oscillators.

%33333333333333333333333333333333333333333333333333333333333333333333333333333333333333333
\section{Linear friction}
\label{sec3}

First we want to verify that our previous small $v$ result, linear in  $v$, is recovered via the energy dissipation. For this situation the $I(\omega_+)$ will not contribute since $\omega_v\rightarrow 0$ and $\omega_+=\omega_1+\omega_2>0$ ($\omega_1,\omega_2>0$). Thus
\begin{equation}
J(\omega_v)=C_- I(\omega_-)=C_- 2\pi\tau\frac{\omega_v^2}{\omega_-}\delta(\omega_-)=H\pi\beta\tau\omega_v^2\delta(\omega_-)
\label{29}
\end{equation}
with $C_-$ given by Eq.~(\ref{19}). Since $\omega_1-\omega_2=\omega_-\rightarrow0$ ($\omega_v\rightarrow 0$) we have  $\sinh{(\beta\hbar\omega_-/2)}\rightarrow\beta\hbar\omega_-/2$, and with $\omega_1=\omega_2=\omega$ we have
\begin{equation}
H=\frac{m^2}{4\sinh^2(\frac{1}{2}\beta m)}\alpha_1\alpha_2, \quad m=\hbar\omega.
\label{30}
\end{equation}
Expression (\ref{29}) can be extended to frequency distributions. As done in Refs.~\cite{hoye12A} and \cite{hoye13} this is obtained by replacing $\alpha_a$ with $\alpha_{Ia}(m_a^2)\, d(m_a^2)$ ($a=1,2$, $m_a=\hbar\omega_a$) which is then integrated. This expression represents the frequency distribution of the oscillators and is obtained from the frequency dependency of the polarizability as given by Eqs.~(46)-(48) of Ref.~\cite{hoye12A} and also by Eqs.~(27)-(29) of Ref.~\cite{hoye13}. This gives the integral
\begin{equation}
\int\delta(\omega_1-\omega_2)\,d(m_1^2)d(m_2^2)
=4m_1^2\hbar^2\,d\omega_1.
\label{31}
\end{equation}
With this extension expression (\ref{29}) becomes ($m_1=m$)
\begin{equation}
J(\omega_v)= 2\tau\omega_v^2 H_0
\label{32}
\end{equation}
\begin{equation}
H_0=\frac{\pi\beta\hbar^2}{2}\int\frac{m^4\alpha_{I1}(m^2)\alpha_{I2}(m^2)}{\sinh^2(\frac{1}{2}\beta m)}\,d\omega.
\label{33}
\end{equation}

The $\omega_v^2$ term in Eq.~(\ref{32}) is the only one that couples to the integration of the interaction $\hat\psi$ in Eq.~(\ref{24}). This interaction is also extended to the dipole-dipole interaction in a straightforward way by replacing $\hat\psi$ with $\hat\psi_{ij}$ ($i,j=1,2,3$). Following Ref.~\cite{hoye13} one for the electrostatic dipole-dipole interaction has
\begin{equation}
\psi_{ij}=-\frac{\partial^2}{\partial x_i\partial x_j}\psi, \quad\psi=\frac{1}{r}
\label{34}
\end{equation}
where here the $\psi$ is the Coulomb interaction between two unit charges (in Gaussian units). Fourier transform in the $x$-and $y$-directions then gives
\begin{equation}
\hat\psi(z_0,k_\perp)=\frac{2\pi e^{-q|z_0|}}{q}
\label{35}
\end{equation}
with $q=k_\perp$ and $k_\perp^2=k_x^2+k_y^2$. Further with $ik_z= q$ for $z> 0$ and  $ik_z=-q$ for $z<0$ ($\partial/\partial x_j\rightarrow-ik_j$)
\begin{equation}
\hat\psi_{ij}(z_0,{\bf k_\perp})=-k_i k_j\hat\psi(z_0,k_\perp).
\label{35b}
\end{equation}
For the integrations in Eq.~(\ref{24}) we need (summation over $i$ and $j$)
\begin{equation}
\hat G(z_0,q)=\hat\psi_{ij}(z_0,{\bf k_\perp})\hat\psi_{ij}(z_0,-{\bf k_\perp})=k_ik_ik_jk_j\hat\psi^2.
\label{36}
\end{equation}
Here some care must be taken in the summations as $ik_z$ follow the sign of $z$ such that
\begin{equation}
-ik_j ik_j=k_x^2+k_y^2+(\pm q)^2=k_\perp^2+q^2=2q^2.
\label{37}
\end{equation}
Thus
\begin{equation}
\hat G(z_0,q)=(2q^2)^2\left(\frac{2\pi e^{-q|z_0|}}{q}\right)^2.
\label{38}
\end{equation}
With this the energy dissipation (\ref{24}) becomes
\begin{equation}
\Delta E=\frac{\rho_1 \rho_2}{(2\pi)^2}\int\limits_{z_1>d, z_2<0}\hat G(z_0,q) J(\omega_v)\,d{\bf k}_\perp dz_1 dz_2.
\label{38a}
\end{equation}
Expression (\ref{38}) is to be integrated together with the $\omega_v^2=(k_x v_x)^2$ of Eq.~(\ref{32}). It is convenient to first perform the $z$-integrations to obtain
\begin{equation}
\hat G(q)=\int\limits_{z_1>d,}\int\limits_{z_2<0}\hat G(z,q)\,dz_1dz_2=(2\pi)^2e^{-2qd}.
\label{39}
\end{equation}
By the further integration one can replace $k_x^2$ with $(k_x^2+k_y^2)/2=q^2/2$ to get the integral ($d{\bf k}_\perp=2\pi q\,dq$)
\begin{equation}
G=\frac{\rho_1\rho_2}{(2\pi)^2}\int\limits_0^\infty\frac{1}{2}q^2\hat G(q)2\pi q\,dq=\frac{3\pi}{8d^4}\rho_1\rho_2.
\label{40}
\end{equation}
Altogether results (\ref{33}) and (\ref{40}) are the same as results (31) and (34) of Ref.~\cite{hoye13}. Accordingly with Eqs.~(\ref{24}), (\ref{28}), and (\ref{32}) the friction force per unit area is also recovered as
\begin{equation}
F=-GvH_0.
\label{41}
\end{equation}

%444444444444444444444444444444444444444444444444444444444444444444444444444444444444444444444444
\section{Friction at zero temperature}
\label{sec4}

The expressions developed above also hold for the general situation with finite velocity. For $T=0$, the result (\ref{41}) is extended in a straightforward way to this latter situation  where now only the $\omega_+$-term contributes since $C_-\rightarrow0$ as $\sinh{(\beta\hbar\omega_i/2)} \rightarrow\infty$ ($i=1,2$).. Thus with Eqs.~(\ref{23}) and (\ref{27}) ($\omega_+=\omega_1+\omega_2$)
\begin{equation}
J(\omega_v)=C_+\pi\tau\frac{\omega_v^2}{\omega_+}[\delta(\omega_+-\omega_v)+\delta(\omega_++\omega_v)]
\label{42}
\end{equation}
where from Eqs.~(\ref{19}) and (\ref{20}) one finds ($\beta\rightarrow\infty$)
\begin{equation}
C_+=\frac{1}{2}\hbar\omega_1\omega_2\alpha_1\alpha_2.
\label{43}
\end{equation}
The two delta functions in Eq.~(\ref{42}) may in the photonic picture be taken to correspond to photons emitted in the same direction as the velocity $\bf v$, or in the direction opposite to it.

Again with frequency distributions the $\alpha_a$ is to be replaced with $\alpha_{Ia}(m_a^2)\,d(m_a^2)$) ($a=1,2$, $m_a=\hbar\omega_a$) to be integrated together with expression (\ref{42}). Thus with $d(m_1^2)d(m_2^2)=4\hbar^2m_1m_2\,d\omega_1d\omega_2$ this gives ($\omega_1+\omega_2=\omega_+=|\omega_v|$)
\begin{equation}
J(\omega_v)=2\pi \tau|\omega_v|\hbar^3\int\limits_0^{|\omega_v|}\omega_1\omega_2m_1m_2\alpha_{I1}(m_1^2)\alpha_{I2}(m_2^2)\, d\omega_1.
\label{44}
\end{equation}

where $\omega_2=|\omega_v|-\omega_1$. To proceed further we need the frequency distribution. As in Ref.~\cite{hoye13} the Drude model for a metal is considered. Its dielectric constant is
\begin{equation}
\varepsilon=1+\frac{\omega_p^2}{\xi(\xi+\nu)}
\label{45}
\end{equation}
where $\xi=i\omega$, $\omega_p$ is the plasma frequency, and $\nu$ represents damping of plasma oscillations due to the finite conductivity of the medium. As shown in Ref.~\cite{hoye13} the low density result can be extended to arbitrary density by replacing polarizability $\alpha=\alpha(\omega)$ with (Eq.~(67) of the reference)
\begin{equation}
2\pi\rho\alpha\rightarrow\frac{\varepsilon-1}{\varepsilon+1}.
\label{46}
\end{equation}
The frequency spectrum follows from the imaginary part of this expression as \cite{hoye13}
\begin{equation}
m^2\alpha_I(m^2)=Dm,\quad D=\frac{\hbar\nu}{\rho (\pi\hbar\omega_p)^2},
\label{47}
\end{equation}
valid for small $m$ ($=\hbar\omega$) far away from the plasma frequency. Now we will assume that both media have the same dielectric constant and have the same density $\rho_1=\rho_2=\rho$. When inserted in expression (\ref{44}) this gives
\begin{equation}
J(\omega_v)=2\pi \tau|\omega_v|\hbar^3 D^2\int\limits_0^{|\omega_v|}\omega_1\omega_2\,d\omega_1=2\tau\omega_v^4 H_P, \quad H_P=\frac{1}{6}\pi\tau\hbar^3 D^2.
\label{48}
\end{equation}
By further integration of $\omega_v^4=(k_x v)^4$ ($v=v_x$) one has $\int k_x^4\,d\phi=k_\perp^4\int_0^{2\pi}\cos^4\phi\,d\phi=2\pi q^4(3/8)$ ($q=k_\perp$) such that $k_x^4$ can be replaced by $3q^4/8$, i.e. we can write
\begin{equation}
J(\omega_v)=2\tau v^4 H_P\cdot\frac{3}{8}q^4.
\label{49}
\end{equation}
Thus in the present case integral (\ref{40}) is replaced by (with $\rho_1=\rho_2=\rho$)
\begin{equation}
G_P=\frac{\rho^2}{(2\pi)^2}\int\limits_0^\infty\frac{3}{8}q^4\hat G(q)2\pi q\,dq=\frac{45\pi}{32d^6}\rho^2.
\label{50}
\end{equation}
With Eqs.~(\ref{24}) and (\ref{36})-(\ref{39}) the dissipated energy follows from Eqs.~(\ref{49}) and (\ref{50}) as
\begin{equation}
\Delta E_P=2\tau H_Pv^4 G_P
\label{51}
\end{equation}
by which the friction force at temperature $T=0$ is
\begin{equation}
F_P=-\frac{\Delta E_P}{2\tau v}=-\frac{15\pi^2}{64d^6}\rho^2 D^2(\hbar v)^3
\label{52}
\end{equation}
with $D$ given by Eq.~(\ref{47}).

%555555555555555555555555555555555555555555555555555555555555555555555555555555555555555555555555
\section{Comparisons with other results}
\label{sec5}

The small $v$ and finite temperature case can also be found explicitly with frequency distribution (\ref{47}). When inserted in expression (\ref{33}) for $H_0$ one recovers result (72) of Ref.~\cite{hoye13} with
\begin{equation}
H_0=\frac{\pi\beta\hbar}{2}D^2\int\limits_0^\infty\frac{m^2\,dm}{\sinh^2(\frac{1}{2}\beta m) }=\frac{2\pi\hbar}{\beta^2}D^2I,
\label{53}
\end{equation}
\begin{equation}
I=\int\limits_0^\infty\frac{x^2 e^{-x}\,dx}{(1-e^{-x})^2}=\sum\limits_{n=1}^\infty\int\limits_0^\infty x^2ne^{-nx}\,dx=2!\sum\limits_{n=1}^\infty\frac{1}{n^2}=\frac{\pi^2}{3},
\label{54}
\end{equation}
and the friction force (\ref{41}) becomes (with $\rho_1=\rho_2=\rho$)
\begin{equation}
F=\frac{3\pi}{8d^4}\rho^2v\frac{2\pi\hbar}{\beta^2}D^2\frac{\pi^2}{3}=\frac{\pi^4}{4\beta^2d^4}\rho^2 D^2\hbar v.
\label{55}
\end{equation}
As found in Ref.~\cite{hoye13} this result, apart from a small factor $\approx 1.2$ is the same as result (97) of Ref.~\cite{volokitin07}. (The explanation for this small difference is probably the extra term in the numerator of Eq.~(92) in Ref.~\cite{volokitin07}~compared to Ref.~\cite{hoye13} where a similar term was neglected.)
The ratio between the friction forces (\ref{55}) and (\ref{52}) is thus
\begin{equation}
\frac{F}{F_P}=\frac{1}{12}\frac{64\pi^2}{5}\left(\frac{d}{\beta\hbar v}\right)^2.
\label{56}
\end{equation}
This ratio deviates by a factor 1/12 from the ratio found in Ref.~\cite{hoye13} when the result  of Pendry was inserted \cite{pendry97}. Pendry's result with dielectric function $\varepsilon=1+i\sigma/(\omega\varepsilon_0)$ is given by Eq.~(62) of Ref.~\cite{pendry97} as
\begin{equation}
F_{\rm Pendry}=\frac{5\hbar\varepsilon_0^2v^3}{2^8\pi^2\sigma^2d^6}\quad (\mbox{for}\quad  v<\frac{d\sigma}{\omega\varepsilon_0}).
\label{57}
\end{equation}
When comparing with our dielectric function (\ref{45}) one sees that $\sigma/\varepsilon_0\rightarrow\omega_p^2/\nu$ (for small $\xi$). (For larger $v$ the $F_{\rm Pendry}$ changes since the frequency spectrum will deviate from expression (\ref{47}) for larger $m$.)

However, in Ref.~\cite{volokitin07} Volokitin and Persson find that the $T=0$ result, here called $F_{\rm VP}$,  should be
\begin{equation}
F_{\rm VP}=6F_{\rm Pendry}.
\label{58}
\end{equation}
(Then the $\sigma/\varepsilon_0$ of Ref.~\cite{pendry97} is the $4\pi\sigma$ of Ref.~\cite{volokitin07}.) The $F_{\rm VP}$ result deviates from our result (\ref{52}) only by a factor 2. We note that the result of Ref.~\cite{pendry97} is based upon its Eq.~(25) which is then used in Ref.~\cite{volokitin07} to obtain $F_{\rm VP}$, not $F_{\rm Pendry}$ for some reason.

%So the latter appears to  contain a misprint. By a closer look we realize that Eq.~(25) of  Ref.~\cite{pendry97} is almost our integrals above put together. The only real difference is that its integration is limited to $k_x>0$. Thus $F_{\rm VP}$ is only one half of our result (\ref{52}). The reason for this difference seems to be that that we also include $k_x<0$. This follows from our Eq.~(\ref{27}) where $\omega_v>0$ contributes in the first $\delta$-function and $\omega_v<0$ in the other (with $\omega_\pm>0$) such that dissipation does not depend upon the sign of $k_x$ relative to $v$.

In Ref.~\cite{pendry10} a general expression for the friction force at $T=0$ is given by its Eq.~(12). There it is mentioned that a factor of two in the original has been corrected. We can compare directly with this expression without limitation to the frequency spectrum (\ref{47}). In general the frequency spectrum is ($m=\hbar\omega$) \cite{hoye13}
\begin{equation}
2\pi\rho\alpha_I(m^2)m^2=-\frac{1}{\pi}\Im R(\omega),\quad R(\omega)=\frac{\varepsilon(\omega)-1}{\varepsilon(\omega)+1}.
\label{63}
\end{equation}
This is inserted in Eq.~(\ref{44}) to rewrite it as
\begin{equation}
J\omega_v)=2\pi\tau|\omega_v|\hbar^3\left(\frac{1}{2\pi^2\rho}\right)^2\frac{1}{\hbar^2}\int\limits_0^{|\omega_v|}\Im R(\omega_1)\Im R(\omega_2)\,d\omega_1.
\label{64}
\end{equation}
The resulting energy dissipated Eq.~(\ref{38a}) then follows by integrating expression (\ref{64}) together with (\ref{39}) multiplied with $\rho^2/(2\pi)^2$ ($\rho_1=\rho_2$) to obtain
\begin{equation}
\Delta E=2\tau v\frac{\hbar}{4\pi^3}\int\limits_{-\infty}^\infty |k_x|\,dk_x\int\limits_{-\infty}^\infty e^{-2qd}\,dk_y\int\limits_{-\infty}^{|\omega_v|}\Im R(\omega_1)\Im R(\omega_2)\,d\omega_1,
\label{65}
\end{equation}
where $\omega_v =k_x v$ and $\omega_2=|\omega_v|-\omega_1$. With friction force $F_x=\Delta E/(2\tau v)$ one will find that our result is two times result (12) of Ref.~\cite{pendry10} (apart from its denominator) where the integration over $k_x$ is limited to $k_x>0$.

Barton has also evaluated the friction force at $T=0$ \cite{barton11B}. This was done via the loss of power ($P=Fv$).  According to Eq.~(6.5) of Ref.~\cite{barton11B} the friction force is found to be (with its $\beta=1$ for metals)
\begin{equation}
F_{\rm B}=12F_{\rm Pendry}
\label{59}
\end{equation}
when a factor $\zeta(5)=1.037$ is disregarded. (This factor may have an explanation similar to the $\zeta(3)\approx1.2$ mentioned below Eq.~(\ref{55}).) When comparing $F_{\rm B}$ with our result (\ref{52}) via Eqs.~(\ref{56})-(\ref{58}) it is seen that $F_{\rm B}=F_P$ and thus coincides with our result. In Ref.~\cite {barton11B} it was remarked below its Eq.~(6.5) that the ratio of disagreement was puzzling. Our findings above give an explanation.

The agreement with the result of Barton may not be unexpected. In Ref.~\cite{hoye11A} we showed the equivalence between the result of time-dependent perturbation theory and the statistical mechanical approach we use, where the Kubo formalism is utilized. In his work Barton uses time-dependent perturbation theory. In Ref.~\cite{hoye11A} we showed equivalence to the result for the pair of osillators considered by Barton in Ref.~\cite{barton10A} at $T=0$.

 Thus we can conclude that our result (\ref{52}) for $F_P$ agrees with the one in Eq.~(\ref{58}), based upon the integral expression given by Pendry (with respect to its $v^3$-dependence for not too high $v$). But its magnitude as found by us is increased by a factor 2. On the other side, when comparing with the force $F_{\rm B}$ from expression (\ref{59}) as found by Barton, also the magnitude of the force agrees.

Pendry also has an expression for the friction force for interacting surface plasma waves with only one frequency, the surface plasma frequency $\omega_{sp}=\omega_p/\sqrt{2}$ \cite{pendry10}. This is expression (11) of the reference. Then a sharp frequency distribution around $\omega_{sp}$ is considered, i.e. Eq.~(\ref{45}) for $\varepsilon$ is considered in the limit $\nu\rightarrow 0$. So with relation (\ref{63}) we get
\begin{equation}
R(\omega)=\frac{\omega_{sp}^2}{\omega_{sp}^2-\omega^2+i\nu\omega}=\frac{\omega_{sp}^2(\omega_{sp}^2-\omega^2-i\nu\omega)}{(\omega_{sp}^2-\omega^2)^2+(\nu\omega)^2}
\label{66}
\end{equation}
\begin{equation}
\Im R(\omega)=-\omega_{sp}^2\pi\delta(\omega^2-\omega_{sp}^2)=-\frac{\pi}{2}\omega_{sp}\delta(\omega-\omega_{sp})
\label{67}
\end{equation}
(for $\omega>0$). The $\delta$-functions follow from $\int_{-\infty}^\infty(a\,dx)/(x^2+a^2)=\pi$ and $\delta(cx)=\delta(x)/c$ ($c=\omega+\omega_{sp}\rightarrow 2\omega_{sp}$).
Apart from sign (using $\xi=-i\omega$ in (\ref{45}) would give plus sign in (\ref{67})) Eq.~(\ref{67}) is only 1/2 of the corresponding expression (14) of Ref.~\cite{pendry10}. This latter expression was used to recover result (11) of the reference on basis of its Eq.~(12). Its Eq.~(11) was first obtained via an independent derivation there. We, however, on basis of expression (\ref{67}) when inserted in Eq.~(\ref{65}) will accordingly find $2\cdot(1/2)^2=1/2$ of result (11) of the reference. So there is a minor discrepancy. On the other side the latter contribution
will be vanishingly small by use of realistic numbers compared with the small $m=\hbar\omega\rightarrow 0$ contribution found above. (I.e. higher frequencies in the frequency distribution, including the resonance peak at the surface plasma frequency, can be neglected.) Barton has performed a similar evaluation with only one frequency to obtain the explicit result (5.26) in Ref.~\cite{barton11B} for the energy dissipated. By inserting numbers one finds, as just mentioned, that this contribution can be neglected. The reason is the exponential factor $\exp{(-4\omega_{sp}}d/v)$. With plasma energy $\hbar\omega_p=9$\,eV for a metal a velocity of $v\approx 2.4\cdot10^6$\,m/s would be needed to have $4\omega_{sp}d/v=1$ even with the tiny spacing of $d=0.1$\,nm.

The presence of quantum friction means that an irreversible process is present where energy is dissipated. This energy should appear as heating of the bodies via the surface waves that are directly involved in the friction. Further transport of the heat will be usual heat conduction not taken into account by the harmonic oscillator model used. Heating also means entropy production which further shows the irreversible nature of the friction problem. This means that after the time-dependent perturbation is turned off, the system has changed and can thus not return to its original state (unless the dissipated energy is removed from it). The condition for this behavior is that the system has started in thermal equilibrium with equal temperatures on both plates, and thermal averages have been evaluated. Thus the method to evaluate friction used here, can not be applied directly to a problem with extension to heat conduction between plates of different temperatures.

\section{Summary}

The main new development of the present work has been to model the relative sliding motion of one of the dielectric plates as a closed loop, therewith facilitating the interpretation of the Casimir friction formalism via the energy dissipated. In this way our formalism previously restricted to the limit of small velocities is extended to more arbitrary velocities. The cases of $T=0$ and finite $T$ are covered by the same formalism. We find the force to be proportional to $v^3$ at $T=0$, and it is in agreement with the result obtained by Barton \cite{barton11B}. Except for a numerical factor 2 (or 12), it is also in accordance with an earlier result of  Pendry \cite{volokitin07,pendry97}. For finite $T$ and small $v$ the force is proportional to $v$ in agreement with a result of Ref.~\cite{volokitin07}.

Our basic method, as in earlier papers, has been to use statistical mechanical methods for harmonic oscillators. That means, applying the Kubo formula to time-dependent cases.

Finally, we mention that the velocity $v$ has been assumed low, in comparison to the velocity of light. Thus our formalism is not intended to describe Casimir friction phenomena under circumstances where the Cherenkov effect comes into play.  Recent papers dealing with the latter kinds of phenomena can be found, for instance,  in Refs. \cite{silveirinha13A} and \cite{silveirinha13B}.

\end{document}